\documentclass[journal]{IEEEtran}
\usepackage[pdftex]{graphicx}
\usepackage{graphicx}
\usepackage{url}
\usepackage{xcolor}
\ifCLASSINFOpdf
\else
\fi

\begin{document}

\title{Enhancing Cancer Prediction in Challenging Screen-Detected Incident Lung Nodules Using Time-Series Deep Learning}
\author{Shahab Aslani, Pavan Alluri, Eyjolfur Gudmundsson, Edward Chandy, John McCabe, Anand Devaraj, Carolyn Horst, Sam M Janes, Rahul Chakkara, Arjun Nair, Daniel C Alexander, SUMMIT consortium, and Joseph Jacob 
\thanks{S. Aslani, E. Gudmundsson, E. Chandy, J. McCabe, S.M. Janes, D.C. Alexander, and J. Jacob are with Centre for Medical Image Computing, Department of Computer Science, Lungs for Living Research Centre, UCL Respiratory, University College London, UK.} 
\thanks{P. Alluri and R. Chakkara are with MANAS AI, London, UK.} 
\thanks{A. Nair is with University College London Hospitals NHS Foundation Trust, London, UK.} 
\thanks{C. Horst is with Lungs For Living Research Centre, UCL Respiratory, University College London, UK.}
\thanks{A. Devaraj is with Royal Brompton and Harefield NHS Foundation Trust, National Heart and Lung Institute, Imperial College, London, UK.} 
\thanks{SUMMIT consortium refers to all co-authors that comprise the SUMMIT study group.}
\thanks{Corresponding author's e-mail: j.jacob@ucl.ac.uk}}
%
%
%
\maketitle

\begin{abstract}
Lung cancer is the leading cause of cancer-related mortality worldwide. Lung cancer screening (LCS) using annual low-dose computed tomography (CT) scanning has been proven to significantly reduce lung cancer mortality by detecting cancerous lung nodules at an earlier stage. Improving risk stratification of malignancy risk in lung nodules can be enhanced using machine/deep learning algorithms. However most existing algorithms: a) have primarily assessed single time-point CT data alone thereby failing to utilize the inherent advantages contained within longitudinal imaging datasets; b) have not integrated into computer models pertinent clinical data that might inform risk prediction; c) have not assessed algorithm performance on the spectrum of nodules that are most challenging for radiologists to interpret and where assistance from analytic tools would be most beneficial. 

Here we show the performance of our time-series deep learning model (DeepCAD-NLM-L) which integrates multi-model information across three longitudinal data domains: nodule-specific, lung-specific, and clinical demographic data. We compared our time-series deep learning model to a) radiologist performance on CTs from the National Lung Screening Trial enriched with the most challenging nodules for diagnosis; b) a nodule management algorithm from a North London LCS study (SUMMIT). Our model demonstrated comparable and complementary performance to radiologists when interpreting challenging lung nodules and showed improved performance (AUC=88\%) against models utilizing single time-point data only. The results emphasise the importance of time-series, multi-modal analysis when interpreting malignancy risk in LCS.

\end{abstract}
\begin{IEEEkeywords}
Computer-Aided Diagnosis, Computed Tomography, Lung Cancer, Longitudinal Study, Deep Learning.
\end{IEEEkeywords}
\IEEEpeerreviewmaketitle

\section{Introduction}
\IEEEPARstart{L}{ung} cancer is the most common cause of cancer death in the world~\cite{WHO}. Early detection of lung cancer using low-dose CT scans in lung cancer screening (LCS) studies allows timely intervention and treatment thereby reducing lung cancer mortality rates. This has resulted in defined lung cancer screening guidelines~\cite{national2011reduced,national2011national,black2014cost,de2020reduced}. The US National Lung Screening Trial (NLST)~\cite{national2011national} and the Dutch-Belgian NELSON Trial~\cite{de2020reduced}, demonstrated overall reductions in lung cancer mortality of at least 20\%~\cite{national2011national,de2020reduced}. The focus of LCS studies is the detection of pulmonary nodules and assessment of nodule growth which may indicate the presence of early lung cancer. However, the majority of screen-detected nodules are either benign or have no bearing on a patient's prognosis. The probability of a lung nodule being malignant is currently determined, in combination with an individual's risk factors, in two ways: (1) at baseline, an assessment of size and associated characteristics (location, density, morphology); and (2) evolution of nodule characteristics - chiefly growth rate - on interval scans. Lung cancer screening studies function optimally when lung nodules are detected with high sensitivity, but, then interpreted with high specificity, to achieve a high discriminatory performance for lung malignancy prediction. Together, this allows the detection of high-risk nodules at an early curable stage. 

Lung cancer screening generates huge volumes of CT imaging that require evaluation by radiologists. Yet in countries such as the UK, there remain national shortages of radiologists to evaluate screening CTs~\cite{royal2021clinical}. The field of lung cancer prediction on CT scans using machine learning and deep learning algorithms has matured following the wide availability of large screening datasets for analysis. Algorithms can now be expertly trained with a breadth of examples of nodule types beyond that which an average radiologist would encounter over an entire career. There is hope that utilizing computer algorithms as objective diagnostic aids for radiologists when interpreting LCS CT imaging may result in faster and better reads of challenging screening cases. 

\textbf{Study Rationale}. In typical radiology workflows, when an abnormality is detected on a CT by a radiologist, prior imaging is sought to better understand how the lesion has changed in appearance over time. Evaluating time-series data to understand disease behaviour is a central tenet of radiology. It would therefore be expected that machines may also improve their performance in predicting malignancy by interrogating time-series data. Yet most previous studies have focused on analysing single timepoint CT datasets ~\cite{liao2019evaluate} acquired as part of lung cancer screening programs~\cite{pearl2014probabilistic}.

Studies examining malignancy risk prediction in LCS have typically indiscriminately assessed all nodules contained within screening cohorts. However, nodules subtypes can vary across the breadth of a screening study. Prevalent cancers seen at the first screen are typically larger and easier to identify by human readers, with or without the aid of computers. Incident cancers however, evolving from pre-existing nodules or de novo, are more challenging to distinguish from benign nodules. Support from computer tools to improve the classification of these challenging nodules would be most beneficial for a radiologist in a screening setting. It is therefore necessary to define the performance of computer-based nodule classification when applied to difficult cases. Lastly, no studies have integrated readily available clinical information with time-series image data when designing an automated system for lung cancer prediction, yet this ignores potentially valuable information that is routinely collected in screening programs. 

Therefore, in our study we propose a deep learning-based model called DeepCAD-NLM-L to improve the prediction of lung cancer likelihood utilizing all available time-series CT data. Our model aggregates both lung-level and nodule-level information thereby leveraging the advantages of both sources of imaging data. We also combine clinical metadata alongside imaging data to aid in lung cancer prediction. 

Anticipating that an integrated time-series model may have utility when classifying the most challenging lung nodules seen in LCS programs, we evaluate classification performance on "small" to "intermediate" sized nodules. We also compare the utility of time-series analyses of CTs using DeepCAD-NLM-L against nodule management algorithms utilised in the SUMMIT LCS study. The contemporary SUMMIT study CTs are higher quality and acquired at lower dose than the CTs contained within NLST. Our subanalysis, therefore, tests the performance of DeepCAD-NLM-L on data out-of-distribution. As part of this analysis, we also examine whether the sensitivity and specificity of human readers versus computer algorithms might provide complementary interpretations of nodule malignancy risk in a LCS setting. 

\textbf{Automated Lung Cancer Detection}. Automated diagnosis of lung cancer using deep learning methods typically addresses either lung-level or nodule-level predictions. For lung-level prediction methods, the entire CT scan is used as an input to the model~\cite{wang2019lung,causey2019lung,jiang2020attentive,gao2019distanced}. Nodule-level lung cancer prediction methods consist of two sequential stages: computer-aided detection (CADe), where the nodule is identified, followed by "diagnosis" (CADd), where a malignancy probability is assigned to the identified nodule and then a cancer/non-cancer label is assigned to the case~\cite{liao2019evaluate,li2020deepseed,khosravan2018s4nd,ding2017accurate, kuan2017deep,liao2019evaluate,trajanovski2018towards,ozdemir20193d,ardila2019end,zhu2018deeplung}.

Nodule-level methods have been shown to be more accurate than lung-level models when predicting malignancy risk in candidate nodules. However, classifying malignancy risk in a nodule remains reliant on the ability of the model to detect the nodule in the first place. Detecting a nodule, in turn, relies on good quality training data. Lung lesions $>$30mm in size, though no longer formally defined radiologically as nodules (as the upper limit in size for a nodule is 30mm), need to be identified by CADe systems as they have a high likelihood of being lung cancer. However, most training datasets do not contain many examples of lesions of this size, thereby representing a potential source of false negatives in automated lung cancer detection systems. On the other hand, lung-level methods are robust to predict the lung cancer when there are big lesions ($>$30mm). Therefore the combination of lung-level and nodule-level methods can boost the lung cancer prediction.

\begin{figure}
  \includegraphics[width=\linewidth]{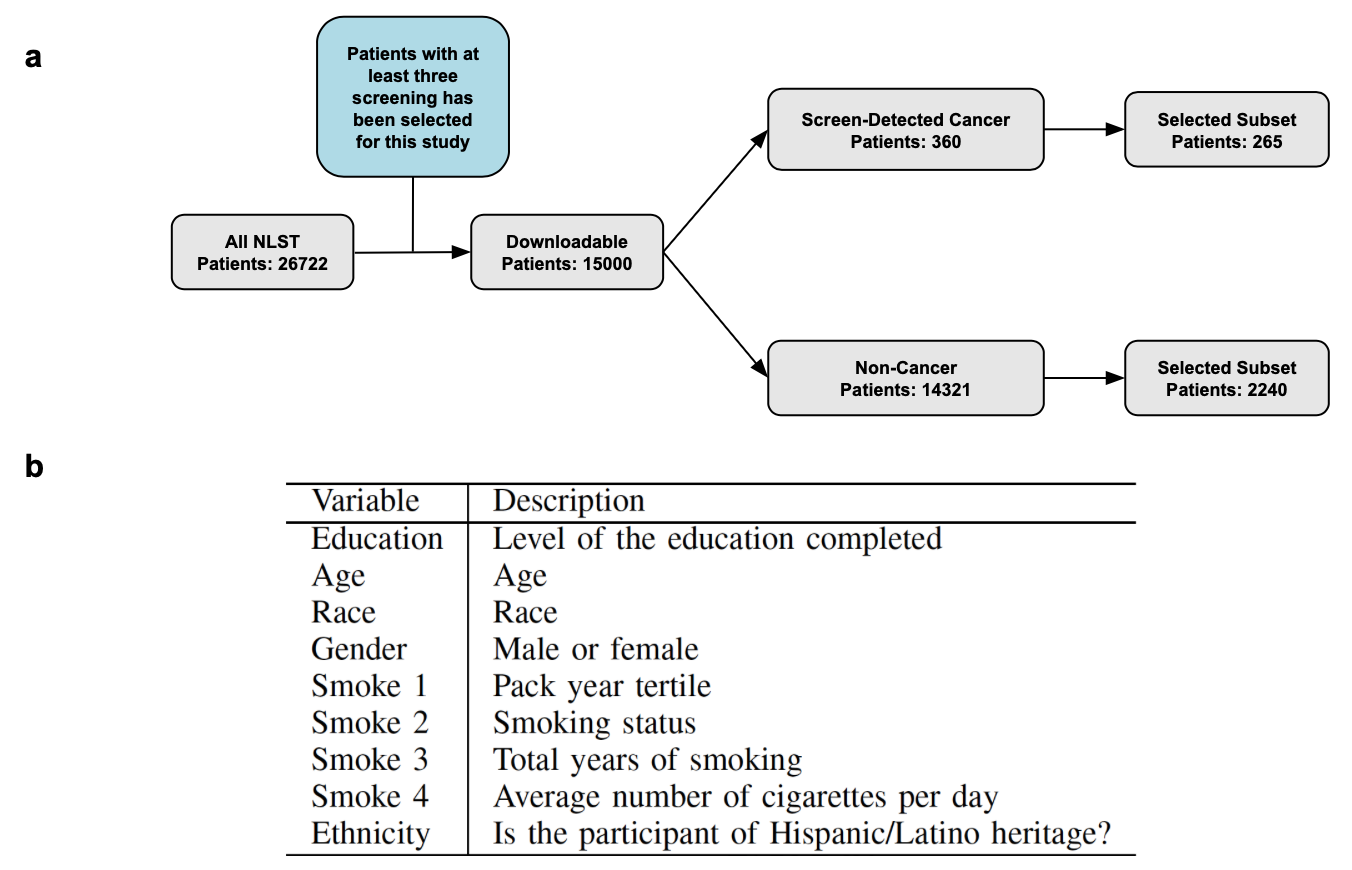}
  \caption{NLST data selection strategy for our study. a) Data from 15,000 NLST patients were available, with up to 3 time point computed tomography scans per patient. All of the available screen-detected incident lung cancers (n=265) in NLST were examined. 2240 patients without cancer were randomly selected for analysis. b) Clinical metadata demographics.}
  \label{fig_data}
  \centering
\end{figure} 

\textbf{Related Work}. Several studies have utilized deep learning-based models for lung cancer prediction. For models considering lung-level predictions, in~\cite{jiang2020attentive} a 3D CNN network has been proposed to predict lung cancer on CT images of the whole lung. Regarding nodule-level prediction methods, Liao et al.,~\cite{liao2019evaluate} proposed an approach that included CADe and CADd systems. They used a 3D Faster R-CNN~\cite{ren2016faster} as the nodule detection network. Later, a set of shallow 3D deep CNNs were used to extract features from candidate nodules and predict the malignancy score using a Leaky Noisy-OR approach~\cite{pearl2014probabilistic}. Recently, Ardila et al.,~\cite{ardila2019end} proposed an approach for lung cancer prediction combining lung-level and nodule-level predictions. The CADe system in this method is a 3D Inception CNN~\cite{szegedy2015going} combined with a Region Proposal Network (RPN)~\cite{ren2016faster} to identify a set of candidate nodules on a participants current and prior CTs. The same 3D Inception CNN was used as a CADd system to calculate the malignancy score. Alternative features from imaging at the lung level were incorporated and boosted performance of the CADd system. They focused on patients where cancer was typically confirmed after the first screening round. Such cancerous nodules are likely to be larger and easier to identify than incident cancers. In~\cite{gao2019distanced} a Long Short-Term Memory Model (LSTM) capable of learning both long-term and short-term dependencies between features was used. A Distanced LSTM allowed evaluation across irregularly sampled intervals though nodule-level features were not studied. Xu et al.,~\cite{xu2019deep} developed a nodule-level prediction method in which the CADd system included a 2D CNN with a Recurrent Neural Network (RNN). This model used four different scans per patient: a baseline CT and CTs one, three and six months later. However, the proposed approach was not fully automated as it did not contain a CADe system. 

To date, no published fully automated method has combined lung-level and nodule-level features with clinical metadata demographics in a longitudinal manner for malignancy estimation.

\begin{figure*}
  \includegraphics[width=\linewidth]{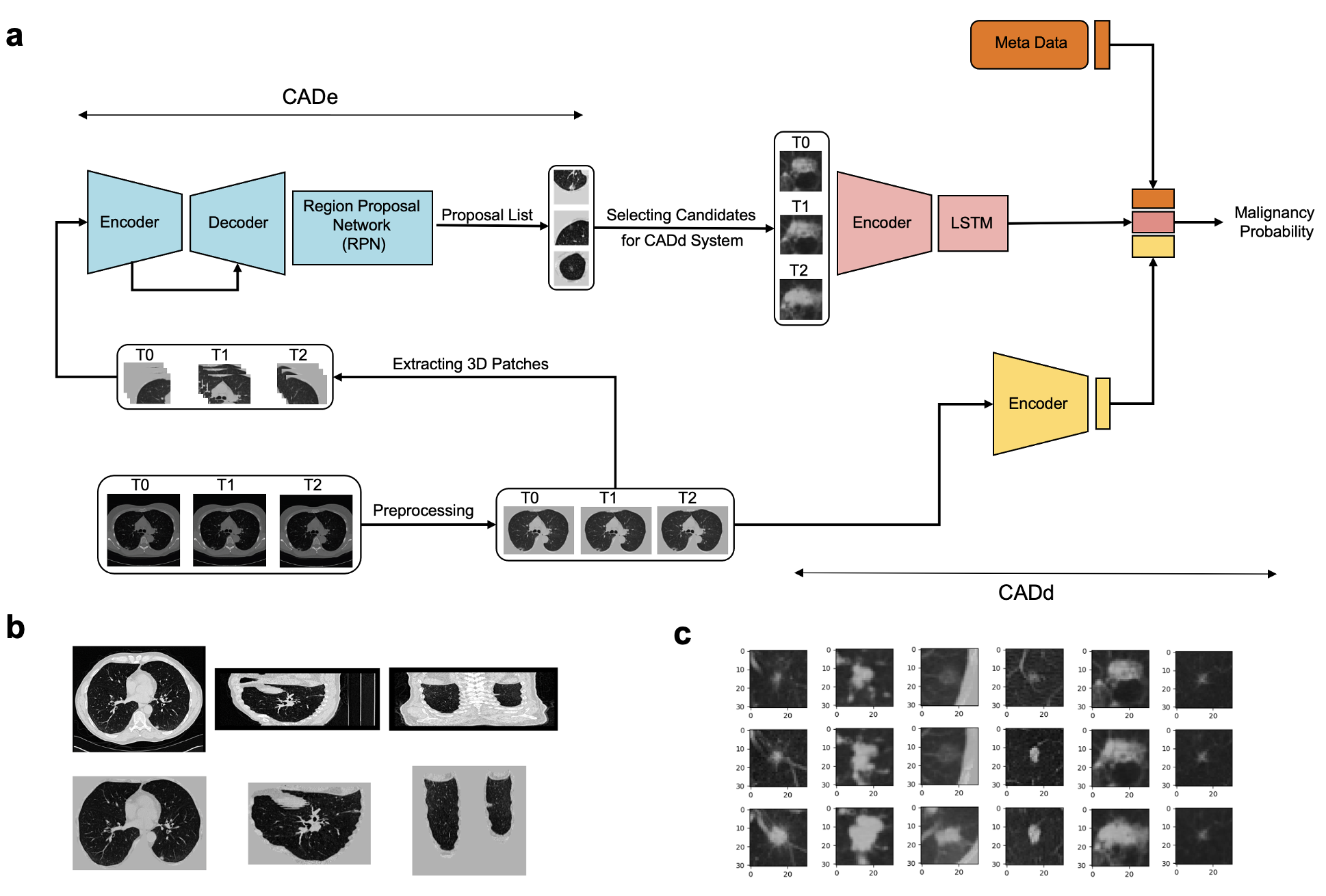}
  \caption{General overview of the proposed DeepCAD-NLM-L method. 
  a) This model includes CADe and CADd systems. In this architecture, the CADe nodule detection system (blue) includes an encoder-decoder followed by a region proposal network. Input to the CADe system is the segmented 3D patches from preprocessed longitudinal scans and its output is a set of nodule proposals. A single nodule that has the highest confidence score is selected as the candidate nodule. Then three 3D patches around the candidate nodule were extracted from all available time-points. Later a CADd system combines the extracted nodule-level (pink Encoder+LSTM) and lung-level (yellow Encoder) longitudinal features with metadata information to predict malignancy. 
 b) Image preprocessing. The top row of images show the original computed tomography scan in axial, sagittal, and coronal planes. Images in the second row show the preprocessed version of the corresponding top row image. 
 c) The output of the patch extraction process for candidate nodules. Each column corresponds to a different time points for a single case. The first, second, and third rows are related to the T0, T1, and T2 time-points, respectfully. }
  \label{fig_method}
\end{figure*}
\section{Materials and Methods}
\label{Method}

\textbf{Datasets}. Our proposed DeepCAD-NLM-L model was validated using two lung cancer screening datasets: The National Lung Screening Trial~\cite{national2011national,national2011reduced} and the SUMMIT Lung Cancer Screening Study~\cite{horst2020delivering}.

NLST was a large randomized multi-center LCS study in the United States in which 26,722 participants underwent three annual screens (T0, T1, and T2) using low-dose CT scans. If cancer was diagnosed on the first CT screen subsequent screening CTs were not performed. Figure~\ref{fig_data}.a displays the NLST cohort analysed in our study. An individual CT was considered cancer-positive if the result of a biopsy or surgical resection was positive during the screening study year. An individual CT was considered cancer-negative if the patient was cancer-free on the follow-up screen~\cite{national2011national,national2011reduced,ardila2019end}. 679 patients had biopsy-confirmed screen detected (n=360) and non-screen detected (n=319) cancers. All NLST CT scans were acquired with the use of multidetector scanners with a minimum of four channels. The acquisition variables were chosen to reduce exposure to an average effective dose of 1.5 mSv as previously described (\url{https://www.nejm.org/doi/full/10.1056/nejmoa1102873}.

The North London-based SUMMIT Lung Cancer Screening Study uses a commercial CADd system to identify lung nodules following which a radiologist accepts or rejects the CAD outputs and performs a second read to ensure that no nodules are missed. The SUMMIT Study aims to assess the implementation of LDCT for lung cancer screening in a high-risk population and to validate a multi-cancer early detection blood test (ClinicalTrials.gov identifier NCT03934866). SUMMIT participants with prevalent cancers (cancers diagnosed on the first screening CT) were excluded from our analysis as we aimed to focus on nodules that were challenging to diagnose. Instead SUMMIT participants with at least two CT images acquired between April 2019-April 2020 were analysed. The nodule management plan in the SUMMIT study required participants with a suspicious nodule on CT (but no lung cancer diagnosis), to be referred to a multi-disciplinary team for further investigation or undergo CTs at 3 and 12 months unless an interim diagnosis of lung cancer was received~\cite{horst2020delivering}.

\textbf{Ethical Statement.}
Approval for SUMMIT retrospective study was obtained from the University College London research ethics committee and Leeds East Research Ethics Committee: 20/YH/0120.

\textbf{Data Availability.}
The NLST dataset in a publicly available dataset and can be requested through the official procedure of the study: \url{https://cdas.cancer.gov/learn/nlst/images/}. The SUMMIT is an on going study and it can not be made publicly available due to confidentiality.

\textbf{Data Preprocessing}. 
All CT scans were converted to Hounsfield Units (HU) (the quantitative scale for describing radiodensity) as per in~\cite{li2020deepseed}. Images were binarized by thresholding at -600HU. Using the 2D/3D connected components and measuring their distances to the center of the image, we extracted lung-connected domains. Erosion and dilation morphological operations were applied to divide the lung mask into right and left lungs. The convex hull of each lung was computed and using a dilation operation masks were combined to create a more accurate binary lung mask. The original image was clipped within the range [-1200 to 600HU] and normalized to [0, 1]. The ultimate mask was used to segment the lung. All voxel values outside the lung were assigned a density of 170HU corresponding to normal tissue density. The image was cropped in all three dimensions to retain just the lung in the image. Finally for each patient, we rigidly registered the later time-point CTs (T2 and T1) to the first time-point CT (T0) using FMRIB's Linear Image Registration Tool (FLIRT)~\cite{jenkinson2001global,jenkinson2002improved}. The result of the preprocessing pipeline on a single NLST CT acquired with different views is shown in Figure~\ref{fig_method}.b.

\textbf{Model Architecture}.
In this section, we introduce our novel fully automated lung cancer prediction model specifically designed to utilise time-series CT data. The proposed model includes a CADe system to detect suspicious nodules and a CADd system to measure a malignancy score. We combined lung-level and metadata features with nodule level features to increase the performance of the CADd system. The general architecture of the proposed method can be seen in Figure~\ref{fig_method}.a. 

Following the idea in~\cite{li2020deepseed}, our CADe nodule detection system includes a 3D U-net~\cite{ronneberger2015u} like Encoder-Decoder for feature extraction followed by a 3D Region Proposal Network~\cite{liao2019evaluate} that enables the model to generate proposals directly. The backbone architecture of the Encoder is a ResNet10~\cite{he2016deep} combined with the mirrored version of the same network with reduced blocks as the Decoder. As suggested in~\cite{li2020deepseed}, the Encoder-Decoder is combined with squeeze-and-excitation blocks~\cite{hu2018squeeze} to generate richer features with more contextual information to detect candidate nodules. Figure~\ref{fig_method}.a shows the overall architecture of our CADe system.

To measure the malignancy score, our proposed CADd system aggregates lung-level, nodule-level, and clinical metadata features extracted from longitudinal data. As a first step, a selection criteria is applied to the outputs of the CADe system that chooses a candidate nodule coordinate and extracts 3D patches from all time-point CTs at the corresponding coordinates (Please refer to section~\ref{Result} for more details). Then a 3D Encoder followed by a long short-term memory (LSTM) layer is used to extract nodule-level features from longitudinal input data. In the second step, the preprocessed 3D scans from all three CT time-points are aggregated to create a single 4D input. A 3D Encoder extracts features from the 4D input corresponding to the lung-level time-series information. In the next step, nodule-level and lung-level features are combined with demographic metadata features to predict the malignancy score. Both the 3D encoders used in the CADd system are ResNet10~\cite{he2016deep}. Figure~\ref{fig_method}.a demonstrates the architecture of our proposed CADd system in detail.

\textbf{Training Implementation Details}. If we consider a set of scans $J$, for each scan, we may have a set of nodules $I$. Each nodule has specific information regarding the volume coordinate $(x_{ij}, y_{ij}, z_{ij})$ and the diameter $(r_{ij})$. To train our CADe system, we needed the coordinate and size of the nodules on the scans. For the NLST dataset, $(x_{ij},r_{ij})$ values representing the axial location and diameter of the nodules were provided. However, $(y_{ij},z_{ij})$ values representing coronal and sagittal locations were missing. Therefore, to prepare the training datasets for our CADe system, we manually annotated 2000 single time-point scans in the NLST dataset. All selected scans have nodules size in the range of 5mm-30mm. In total, 3081 nodule locations were labeled (1-2 nodules per scan, on average) using ITK-SNAP software~\cite{py06nimg}. For training the CADe system, we divided the whole annotated dataset into training (70\%), validation (15\%) and test (15\%) sets. Cases were specifically selected with respect to nodule size to ensure a wide distrubtion of nodule size in each dataset. As the input of the CADe system, we extracted random 3D patches of $128\times128\times128$ from preprocessed scans followed by additional data augmentation. We trained our model using the stochastic gradient descent optimizer with an initial learning rate of 0.001. The batch size was fixed at 8 and the maximum number of training epochs was 100 for all experiments. Focal loss function~\cite{lin2017focal} was used to train the model for capturing more true positives amongst all nodule candidates.

To train the proposed CADd model, as a first step, for each patient we extract clinical metadata features comprising 9 variables (Figure~\ref{fig_data}.b). We studied the whole clinical metadata features to select the most informative attributes for lung cancer prediction. For this purpose, we implemented feature selection algorithm using Random Forest (RF) classifier. As a result of this process, we selected 9 features including: level of education completed, pack-year tertile, age, race, gender, ethnicity, smoking status at T0, total years of smoking, and the average number of cigarettes smoked per day. 

In the second step, an independent 3D encoder (ResNet10) is designed to predict malignancy scores and extract features on time-series data based on lung-level information. For the input to the model, we concatenate all available preprocessed scans (T0, T1, and T2) of the patient as a 4D input. This model was trained using a stochastic gradient descent optimizer with an initial learning rate of 0.001, batch size of 8, and cross-entropy as the loss function. After finalizing the training procedure, we treat the model as a feature extractor by removing the last fully connected layer and extract a feature vector including 512 nodes from the output of penultimate fully connected layer. We categorise these 512 features as lung-level time-series features.

In the last step, we established a 3D encoder (ResNet10) combined with additional layers including an LSTM and two fully connected layers. Input to this model is the nodule level information achieved from the CADe system. The output of the CADe system is a set of detected nodules $(x_{ij},y_{ij},z_{ij},r_{ij},p_{ij})$, where $(x_{ij},y_{ij},z_{ij})$ is the central coordinate, $r_{ij}$ is the radius, and $p_{ij}$ is the confidence score of a nodule $i$ in scan $j$. We first extract this information related to detected nodules on the latest available time-point CT. Then from the set of potential nodules, a single candidate nodule that has the highest confidence score is selected. According to the central coordinate information of the selected candidate nodule, we extract three 3D patches of $61\times61\times61$ around the nodule from all available time-points. An example of extracted patches from all three time-points for the selected candidate nodule can be seen in Figure~\ref{fig_method}.c. We use these patches as inputs to the model. To get the malignancy score we add 512 features (lung-level) and 9 clinical features (metadata) extracted in steps 1 and 2 to the last fully connected layer of the model. We train the proposed model using a stochastic gradient descent optimizer with an initial learning rate of 0.001, batch size of 16, and cross-entropy as the loss function. The proposed model was implemented in Python language\footnote{\url{https://www.python.org}} using Pytorch~\cite{NEURIPS2019_9015}. All experiments were done on a Nvidia Titan RTX 24GB GPU.

\textbf{Statistics}. Five statistical measures were used to evaluate and compare different versions of the proposed DeepCAD model: Accuracy (ACC), Sensitivity (SE), Specificity (SE), F1-Score (F1), Area Under Receiver Operating Characteristic Curve (AUC). In all experiments 5-fold cross-validation was implemented on the training sets and the results on the test sets are expressed as the average of 5-fold cross-validation.
\section{Experimental Analysis and Results}
\label{Result}

\textbf{Evaluation on NLST dataset}. To measure the performance of the proposed DeepCAD-NLM-L, three different experiments were implemented on the NLST dataset. 

In the first experiment, the 265 cancer cases (Figure~\ref{fig_data}.a) were balanced with 265 cancer-negative cases. Our analyses used single features or combinations of the three different features (nodule-level, lung-level and clinical metadata), in ablation studies of the proposed model (Table~\ref{tab4}). To examine the benefits in using longitudinal datasets, we compared the DeepCAD-NLM-L (longitudinal) model with a DeepCAD-NLM-S (single time point) model trained and tested on the latest timepoint CT (T2) only [i.e. a DeepCAD model which omitted the LSTM layer (refer to section~\ref{Method})]. We also compared DeepCAD-NLM-L to malignancy classification using traditional machine learning methods such as a Support Vector Machine (SVM) and Random Forest Classifier (RFC) on clinical metadata and with the Brock University clinical lung cancer risk calculator (PanCan)~\cite{tammemagi2013selection}.

\begin{figure*}
    \centering
    \includegraphics[width=180mm]{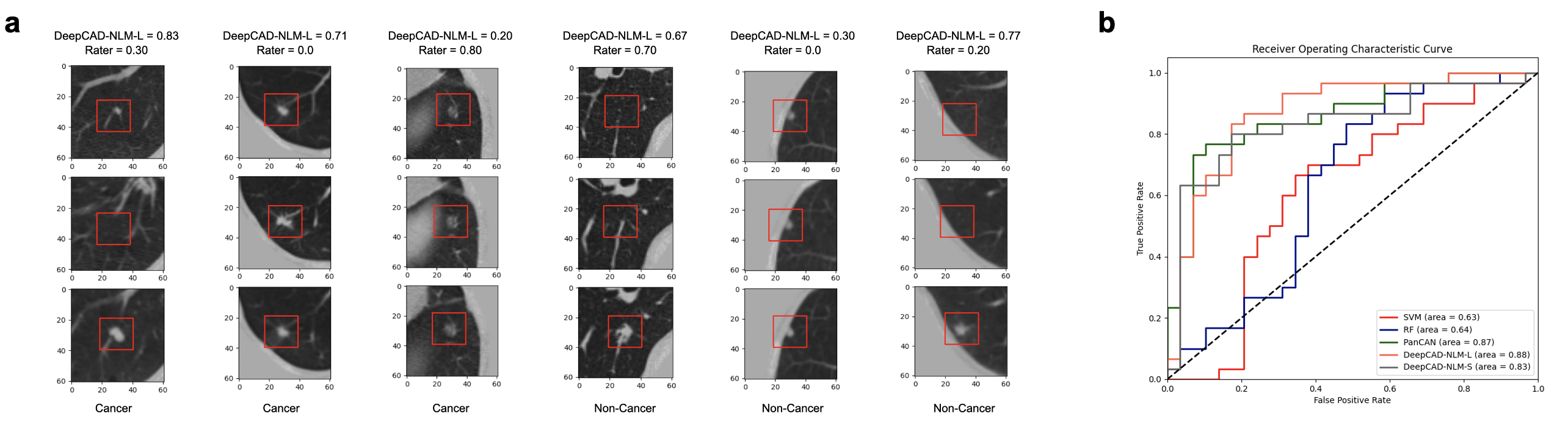}
    \caption{Some results for NLST dataset. a) Output classification results for malignancy prediction for six different patients in the NLST dataset. Each column corresponds to a patient; each row relates to a separate CT time-point (from top to bottom: T0, T1, and T2). The red bounding box contains the candidate nodule. The malignancy prediction output of the proposed model and rater (average of rater1 and rater2) is shown above the top image. The true nodule diagnosis is shown under the bottom image. b) Comparison of the AUC curve of DeepCAD-NLM-L with other approaches in experiment 1 using the NLST dataset.}
    \label{fig_modelout}
\end{figure*}

Table~\ref{tab4} shows the results of experiment 1 on the NLST dataset. The DeepCAD-NLM-L method combining all available features (nodule level, lung level and clinical metadata) and time-points (T0, T1 and T2) outperformed all other models in sensitivity (0.84) with little compromise in specificity (0.87), resulting in the highest AUC (0.88) (Figure~\ref{fig_modelout}.b).

In the second experiment using NLST data, we aimed to compare the performance of the DeepCAD-NLM-L model with that of two thoracic radiologists, on a subset of challenging incident screen-detected nodules. The challenging nodules had previously been part of the DeepCAD-NLM-L model training dataset. Following removal of the challenging nodules from the training dataset (give details of the nodule composition), DeepCAD-NLM-L was retrained on the remaining 85 cancer cases and 180 additional non-cancer cases. Instead of aiming to simply outperform a radiologist in cancer prediction, we hypothesized that interpretation of difficult nodules by DeepCAD-NLM-L and radiologists would show a complementary sensitivity and specificity valuable for LCS studies. The selected nodules in the test dataset included 25 cancer cases and 75 non-cancer cases. Only nodules between 5mm-10mm in size at the first CT time-point were considered. Cancer and non-cancer cases were equally divided into spiculated and non-spiculated nodules as defined by the NLST study data. All CTs had a maximal slice thickness of 2.5mm and all three CT timepoints were analysed. Comparison with human performance was achieved by having two radiologists each with an average of 10 years clinical experience independently assess each set of scans in the test dataset. To simulate clinical practice, for each patient, the radiologists read the longitudinal series of scans side-by-side, and assigned a per-patient diagnosis of cancer or non-cancer, together with measure of diagnostic certainty expressed as a percentage. Both the DeepCAD-NLM-L longitudinal models and the radiologists read the scans in two ways: first, using all three timepoints, and second, using only the first and last timepoints (T0, T1 and T2). To avoid recall bias, the radiologists performed the reading exercise four weeks apart. Whilst the timepoints of the CTs were known to the radiologists, the scans were read in a random order. 

Table~\ref{tab5} summarizes the results of Experiment 2 on the NLST dataset. DeepCAD-NLM-L showed good sensitivity (0.92) whilst the radiologists showed good specificity (0.72-0.77). The PanCan model demonstrated excellent specificity (0.98). The findings suggest that an initial analysis of screening CTs with DeepCAD-NLM-L followed by evaluation of concerning nodules with the PanCan model might optimise nodule management in LCS studies. Furthermore, the similar performances of both the radiologists and DeepCAD-NLM-L when assessing two as opposed to three time-point CTs, suggests that an intermediate annual time-point does not influence malignancy prediction. Figure~\ref{fig_modelout}.a depicts the malignancy probability classification of DeepCAD-NLM-L and the radiologists for six subjects.

In experiment 3, we tested DeepCAD-NLM-L on 2240 NLST cases (265 cancer, 1975 non-cancer) to evaluate model performance on a sample size more representative of lung cancer screening programs. Performance, expressed as the average of 5 cross validation folds on the test datasets was: 72\%, 83\%, 72\% and \%84, for accuracy, sensitivity, specificity and AUC, respectively.

\begin{table*}[]
\centering
\begin{tabular}{llllllll}
\hline
Method                 & Features & CT Time-points  & ACC  & SE   & SP   & F1   & AUC \\ \hline
Support Vector Machine~\cite{cristianini2000introduction} & Metadata & -        & 0.63 & 0.57 & 0.70 & 0.61 & 0.63     \\
Random Forest~\cite{liaw2002classification} & Metadata & -      & 0.64 & 0.69 & 0.60 & 0.66 & 0.64     \\
PanCan~\cite{tammemagi2013selection} & Metadata & - & 0.74 & 0.49 & 0.98 & 0.66 & 0.87     \\
DeepCAD-NLM-S                & Nodule + Lung + Metadata & T2    & 0.75 & 0.73 & 0.77 & 0.75 & 0.83     \\
DeepCAD-N-L                & Nodule & T0, T1, T2         & 0.72 & 0.87 & 0.57 & 0.75 & 0.71     \\
DeepCAD-L-L                & Lung   & T0, T1, T2        & 0.78 & 0.64 & 0.94 & 0.75 & 0.81     \\
DeepCAD-NLM-L                & Nodule + Lung + Metadata & T0, T1, T2    & 0.85 & 0.84 & 0.87 & 0.85 & 0.88     \\ 
\hline
\end{tabular}
\caption{Performance of the different models using varying features in Experiment 1 of the NLST dataset.}
\label{tab4}
\end{table*}

\begin{table*}[]
\centering
\begin{tabular}{llllllll}
\hline
Method  & Features& CT Time-points &  ACC & SE & SP & F1 & AUC \\ \hline
Rater 1 & - & T0, T1, T2 & 0.72 & 0.72 & 0.72 & 0.56 & 0.78     \\
Rater 2 & - & T0, T1, T2 & 0.74 & 0.60 & 0.77 & 0.53 & 0.75     \\
Rater 1 & - & T0, T2 & 0.71 & 0.72 & 0.71 & 0.55 & 0.76     \\
Rater 2 & - & T0, T2 & 0.74 & 0.64 & 0.77 & 0.55 & 0.76     \\
PanCan~\cite{tammemagi2013selection} & Metadata & - & 0.79 & 0.20 & 0.98 & 0.32 & 0.80     \\
DeepCAD-NLM-L & Nodule + Lung + Metadata & T0, T1, T2 & 0.71  & 0.92  & 0.64 & 0.61 & 0.77     \\
DeepCAD-NLM-L & Nodule + Lung + Metadata & T0, T2 & 0.71 & 0.92 & 0.64 & 0.61 & 0.76     \\
\hline
\end{tabular}
\caption{Performance of the proposed DeepCAD method and two radiologists in the Experiment 2 on the NLST dataset.}
\label{tab5}
\end{table*}

\begin{table}[]
\centering
\begin{tabular}{llllll}
\hline
Method &  ACC & SE & SP & F1 & AUC \\ \hline
SUMMIT Outcome & 0.93 & 0.66 & 1.00 & 0.80 & -     \\
PanCan~\cite{tammemagi2013selection} & 0.73 & 0.77 & 0.71 & 0.53 & 0.82      \\
DeepCAD-NLM-L     & 0.75 & 0.83 & 0.73 & 0.58 & 0.80     \\
\hline
\end{tabular}
\caption{Comparison of DeepCAD with radiologist performance on the SUMMIT dataset.}
\label{tab6_summit}
\end{table}

\textbf{Evaluation on SUMMIT dataset}. To evaluate DeepCAD-NLM-L on contemporaneous lung cancer screening imaging, we analysed data from participants in the SUMMIT LCS study. The NLST data used to train DeepCAD-NLM-L was acquired approximately 20 years ago. Interval improvements in imaging have meant that todays CTs have a much narrower slice thickness ($<$1mm in SUMMIT vs 2.5mm typically seen in NLST) and are performed at low-dose using iterative reconstruction techniques compared to filtered back projection CTs at standard dose in NLST.  We tested the DeepCAD-NLM-L model used in experiment 1 of the NLST dataset on a subset of the SUMMIT dataset. To further challenge DeepCAD with out-of-distribution data, cases with varying time intervals between CTs were evaluated. The selected subset is 89 consecutive cases with two or more timepoint CTs included: baseline and 3 month follow up CTs (n=30), baseline and 12 month follow up CTs (n=33); baseline, 3 month and 12 month follow up CTs (n=26). The performance of the DeepCAD-NLM-L model was compared with that of the SUMMIT nodule management algorithm to enable an understanding of the practical impact of a deep learning system. For the purposes of this comparison, a scan was given a label of "cancer" when the SUMMIT radiologist had indicated an urgent patient referral for a suspected lung cancer was required (according to the SUMMIT nodule management protocol).

Table~\ref{tab6_summit} shows the performance of DeepCAD-NLM-L and the SUMMIT nodule management algorithm on the SUMMIT dataset. DeepCAD-NLM-L showed good sensitivity (0.83) despite being trained on out-of-distribution NLST data. Radiologists applying the SUMMIT nodule management algorithm achieved excellent specificity when compared to the DeepCAD-NLM-L model. Taken together, these results suggest that such a computer algorithm could be used to first enhance sensitivity, following which a radiologist could provide high specificity to enable optimal combined performance for LCS. The results indicate the advantages that could be gained in LCS programs when computer algorithms are combined with either nodule management algorithms or cancer risk calculators.
\section{Discussion and Conclusion}
In this paper, we propose a fully automated pipeline for lung cancer prediction from CT scans. Our DeepCAD-NLM-L model encompasses a nodule detection and malignancy prediction system that combines lung-level, nodule-level, and clinical metadata information to increase prediction performance. Importantly, by leveraging valuable information contained within time-series CT data, our model achieves improved prediction of the likelihood of lung cancer in the most challenging lung nodule subtypes. Our model also demonstrates complementary performance when compared to radiologist interpretation of incident lung nodules and cancer risk predictors emphasising the importance of integrating human and computer intelligence in LCS programs.

The diagnosis of lung cancer on imaging by radiologists has evolved iteratively over the past 100 years. For mid- or late-stage lung cancer a confident diagnosis can be made on single timepoint imaging. However as the possibility of stage-shifting cancer diagnosis with earlier detection has become apparent, radiologists have focused on studying changes in morphology of smaller nodules over time to better distinguish benign from malignant lesions. Longitudinal nodule evaluation underpins lung cancer screening programs, and is essential to reduce morbidity in LCS studies from unnecessary investigation of benign lesions and the avoidance of missing a cancer diagnosis.

However to date, most computer algorithms assessing malignancy risk in lung nodules only study single timepoint imaging. Just as radiologists would be remiss if ignoring pertinent historic patient information in the form of old imaging, when evaluating lung nodules, intuitively, one would imagine that computer algorithms would improve malignancy estimation of early cancers by considering any available longitudinal imaging. This concept formed the central premise of our study and was confirmed in the finding that a DeepCAD-NLM-L model evalauting three time-point CTs demonstrated better performance than the same model that only utilsied a single time-point CT (Table II). The benefits of using a time-series model for cancer prediction were also emphasized in results obtained when DeepCAD-NLM-L assessed SUMMIT study data that was inherently different in composition (image quality and reconstruction, radiation dose and CT time intervals) to that used to train DeepCAD-NLM-L.

Our analyses have also highlighted the limitations that result when focusing solely on either nodule level or lung level features. The DeepCAD-N-L model that only considered nodule level features had the highest sensitivity (87\%) of all the models in Experiment 1 but had limited specificity (57\%). The DeepCAD-L-L model that utilised lung-level features conversely had the highest specificity (94\%) of all the DeepCAD models at the expense of sensitivity (64\%). Combining lung-level and nodule-level features would appear to optimise the necessary trade-off between sensitivity and specificity confirmed in DeepCAD-NLM-L having the highest F1 measure, (85\%) of all models (Table II).  

Prior studies evaluating lung nodules have generally had a "blunderbuss" approach to nodule datasets, by focusing on all nodules, rather than indeterminate and challenging cases. In doing so, the performance of such models is potentially artificially inflated by the simultaneous inclusion of both easily dismissed nodules (nodules that are too small and would not result in any meaningful intervention even if classified as cancer at an earlier timepoint) and nodules that are clearly cancerous. The lung cancer prediction model in \cite{ardila2019end} reported AUC values of 94.4\% for prevalent  cancers and 87.3\% (sensitivity=64.7\%; specificity=95.2\%) for cases where cancer was diagnosed in the first two years of screening. DeepCAD-NLM-L demonstrated comparable performance metrics to \cite{ardila2019end} when evaluating incident nodule's alone (AUC=88\%; sensitivity=84\%; specificity=87\%) with a high resultant F1 score (85\%). 

Yet the acute need when interpreting nodules in lung cancer screening cohorts is assistance in the confident characterisation of incident cancers thereby minimising false positive and negative reporting. When we used DeepCAD-NLM-L to assess a carefully curated subset of spiculated and non-spiculated nodules of indeterminate size (5-10mm) which are exactly the types of nodules that consume a disproportionate amount of radiologist interpretation time, DeepCAD-NLM-L correctly identified cancerous nodules when present on longitudinal CTs. One would expect the discriminatory ability of the DeepCAD-NLM-L model to distinguish cancers from non-cancers to be markedly improved compared to single time-point trained models (DeepCAD-NLM-S); in other words, DeepCAD-NLM-L should have a high specificity. Such a high specificity would in turn allow automated triage of scans with a high probability of lung cancer to urgent lung cancer referral. Interestingly, in our experiments this proved to be only partially true. While the DeepCAD-NLM-L model outperformed its single time-point (DeepCAD-NLM-S) and traditional machine-learning model counterparts (SVM and RF), it could not match the performance of a validated risk prediction score (the Brock/PanCan model) or individual radiologists on the NLST dataset, nor a rigorous nodule management protocol on the SUMMIT dataset. Conversely radiologists showed good specificity suggesting that a composite approach whereby DeepCAD-NLM-L pre-reads time-series CTs and highlights nodules of concern for definitive evaluation by a radiologist, could represent an effective screening workflow. With some lung cancer screening programs considering imaging at 2-yearly intervals, it was reassuring to note that DeepCAD-NLM-L with two CT time-points performance was maintained when the second time-point CT (T1) was omitted from time-series analyses in experiment 2 (Table III).  

Our study had several limitations. In our proposed pipeline, one candidate nodule was specified for analysis from all the potential nodules generated by the CADe system. This could constrain our CADd system in situations when the model selects a noncancerous nodule as the candidate nodule in cases containing other cancerous nodules. To mitigate this problem, we aim to develop a new pipeline that uses a selection of candidate nodules as input to the CADd system.

DeepCAD-NLM-L also focuses on the detection of lung nodules which is not synonymous with lung cancer detection. Lung cancers can be located in regions other than the lung itself. An important example of this is small cell lung cancer which may be entirely contained within the mediastinum and therefore not detected by lung nodule detection systems. A future aim would be to incorporate information from mediastinal lung reconstruction kernals when assessing patient level cancer risk.

In conclusion, in this paper, we show that the combination of different levels of features in the DeepCAD-NLM-L model including clinical metadata and imaging data at the lung and nodule-level across longitudinal time-point CTs provides a good estimation of malignancy particularly for incident screening nodules that are challenging for a radiologist to interpret. DeepCAD-NLM-L shows complementary performance metrics of sensitivity and specificity when compared to nodule management algorithms and nodule risk estimators emphasizing the role such tools may have in rationalising the assessment of lung cancer screening CTs.
\section*{Acknowledgments}
This work was supported by Cancer Research UK (C68622/A29390). The authors thank the National Cancer Institute for access to NCI's data collected by National Lung Screening Trial (NLST). The statement contained herein are solely those of the authors and do not represent or imply concurrence or endorsement by NCI. JJ and this research was supported by Wellcome Trust Clinical Research Career Development Fellowship 209553/Z/17/Z. For the purpose of open access, the author has applied a CC-BY public copyright licence to any author accepted manuscript version arising from this submission. This project, JJ, EG, SA, SMJ, AN and DCA were also supported by the NIHR UCLH Biomedical Research Centre, UK.
\section*{Conflicts of Interest}
JJ reports fees from Boehringer Ingelheim, Roche,Takeda, NHSX and GlaxoSmithKline unrelated to the submitted work. JJ was supported by Wellcome Trust Clinical Research Career Development Fellowship 209,553/Z/17/Z. SMJ reports fees from Astra-Zeneca, Bard1 Bioscience, Achilles Therapeutics, and Jansen unrelated to the submitted work. SMJ received assistance for travel to meetings from Astra Zeneca to American Thoracic Conference 2018 and from Takeda to World Conference Lung Cancer 2019 and is the Investigator Lead on grants from GRAIL Inc, GlaxoSmithKline plc and Owlstone.  
\ifCLASSOPTIONcaptionsoff
  \newpage
\fi
\bibliographystyle{IEEEtran}
\bibliography{ref}

\begin{thebibliography}{10}
\providecommand{\url}[1]{#1}
\csname url@samestyle\endcsname
\providecommand{\newblock}{\relax}
\providecommand{\bibinfo}[2]{#2}
\providecommand{\BIBentrySTDinterwordspacing}{\spaceskip=0pt\relax}
\providecommand{\BIBentryALTinterwordstretchfactor}{4}
\providecommand{\BIBentryALTinterwordspacing}{\spaceskip=\fontdimen2\font plus
\BIBentryALTinterwordstretchfactor\fontdimen3\font minus
  \fontdimen4\font\relax}
\providecommand{\BIBforeignlanguage}[2]{{%
\expandafter\ifx\csname l@#1\endcsname\relax
\typeout{** WARNING: IEEEtran.bst: No hyphenation pattern has been}%
\typeout{** loaded for the language `#1'. Using the pattern for}%
\typeout{** the default language instead.}%
\else
\language=\csname l@#1\endcsname
\fi
#2}}
\providecommand{\BIBdecl}{\relax}
\BIBdecl

\bibitem{WHO}
W.~H. Organization, ``A report about cancer,'' p. ONLINE, 2021.

\bibitem{national2011reduced}
N.~L. S. T.~R. Team, ``Reduced lung-cancer mortality with low-dose computed
  tomographic screening,'' \emph{New England Journal of Medicine}, vol. 365,
  no.~5, pp. 395--409, 2011.

\bibitem{national2011national}
NLST, ``The national lung screening trial: overview and study design,''
  \emph{Radiology}, vol. 258, no.~1, pp. 243--253, 2011.

\bibitem{black2014cost}
W.~C. Black, I.~F. Gareen, S.~S. Soneji, J.~D. Sicks, E.~B. Keeler, D.~R.
  Aberle, A.~Naeim, T.~R. Church, G.~A. Silvestri, J.~Gorelick \emph{et~al.},
  ``Cost-effectiveness of ct screening in the national lung screening trial,''
  \emph{N Engl J Med}, vol. 371, pp. 1793--1802, 2014.

\bibitem{de2020reduced}
H.~J. de~Koning, C.~M. van~der Aalst, P.~A. de~Jong, E.~T. Scholten,
  K.~Nackaerts, M.~A. Heuvelmans, J.-W.~J. Lammers, C.~Weenink, U.~Yousaf-Khan,
  N.~Horeweg \emph{et~al.}, ``Reduced lung-cancer mortality with volume ct
  screening in a randomized trial,'' \emph{New England Journal of Medicine},
  vol. 382, no.~6, pp. 503--513, 2020.

\bibitem{royal2021clinical}
R.~C. of~Radiologists, ``Clinical radiology uk workforce census 2021 report,''
  2021.

\bibitem{liao2019evaluate}
F.~Liao, M.~Liang, Z.~Li, X.~Hu, and S.~Song, ``Evaluate the malignancy of
  pulmonary nodules using the 3-d deep leaky noisy-or network,'' \emph{IEEE
  transactions on neural networks and learning systems}, vol.~30, no.~11, pp.
  3484--3495, 2019.

\bibitem{pearl2014probabilistic}
J.~Pearl, \emph{Probabilistic reasoning in intelligent systems: networks of
  plausible inference}.\hskip 1em plus 0.5em minus 0.4em\relax Elsevier, 2014.

\bibitem{wang2019lung}
J.~Wang, R.~Gao, Y.~Huo, S.~Bao, Y.~Xiong, S.~L. Antic, T.~J. Osterman, P.~P.
  Massion, and B.~A. Landman, ``Lung cancer detection using co-learning from
  chest ct images and clinical demographics,'' in \emph{Medical Imaging 2019:
  Image Processing}, vol. 10949.\hskip 1em plus 0.5em minus 0.4em\relax
  International Society for Optics and Photonics, 2019, p. 109491G.

\bibitem{causey2019lung}
J.~L. Causey, Y.~Guan, W.~Dong, K.~Walker, J.~A. Qualls, F.~Prior, and
  X.~Huang, ``Lung cancer screening with low-dose ct scans using a deep
  learning approach,'' \emph{arXiv preprint arXiv:1906.00240}, 2019.

\bibitem{jiang2020attentive}
H.~Jiang, F.~Gao, X.~Xu, F.~Huang, and S.~Zhu, ``Attentive and ensemble 3d dual
  path networks for pulmonary nodules classification,'' \emph{Neurocomputing},
  vol. 398, pp. 422--430, 2020.

\bibitem{gao2019distanced}
R.~Gao, Y.~Huo, S.~Bao, Y.~Tang, S.~L. Antic, E.~S. Epstein, A.~B. Balar,
  S.~Deppen, A.~B. Paulson, K.~L. Sandler \emph{et~al.}, ``Distanced lstm:
  time-distanced gates in long short-term memory models for lung cancer
  detection,'' in \emph{International Workshop on Machine Learning in Medical
  Imaging}.\hskip 1em plus 0.5em minus 0.4em\relax Springer, 2019, pp.
  310--318.

\bibitem{li2020deepseed}
Y.~Li and Y.~Fan, ``Deepseed: 3d squeeze-and-excitation encoder-decoder
  convolutional neural networks for pulmonary nodule detection,'' in \emph{2020
  IEEE 17th International Symposium on Biomedical Imaging (ISBI)}.\hskip 1em
  plus 0.5em minus 0.4em\relax IEEE, 2020, pp. 1866--1869.

\bibitem{khosravan2018s4nd}
N.~Khosravan and U.~Bagci, ``S4nd: Single-shot single-scale lung nodule
  detection,'' in \emph{International Conference on Medical Image Computing and
  Computer-Assisted Intervention}.\hskip 1em plus 0.5em minus 0.4em\relax
  Springer, 2018, pp. 794--802.

\bibitem{ding2017accurate}
J.~Ding, A.~Li, Z.~Hu, and L.~Wang, ``Accurate pulmonary nodule detection in
  computed tomography images using deep convolutional neural networks,'' in
  \emph{International Conference on Medical Image Computing and
  Computer-Assisted Intervention}.\hskip 1em plus 0.5em minus 0.4em\relax
  Springer, 2017, pp. 559--567.

\bibitem{kuan2017deep}
K.~Kuan, M.~Ravaut, G.~Manek, H.~Chen, J.~Lin, B.~Nazir, C.~Chen, T.~C. Howe,
  Z.~Zeng, and V.~Chandrasekhar, ``Deep learning for lung cancer detection:
  tackling the kaggle data science bowl 2017 challenge,'' \emph{arXiv preprint
  arXiv:1705.09435}, 2017.

\bibitem{trajanovski2018towards}
S.~Trajanovski, D.~Mavroeidis, C.~L. Swisher, B.~G. Gebre, B.~S. Veeling,
  R.~Wiemker, T.~Klinder, A.~Tahmasebi, S.~M. Regis, C.~Wald \emph{et~al.},
  ``Towards radiologist-level cancer risk assessment in ct lung screening using
  deep learning,'' \emph{arXiv preprint arXiv:1804.01901}, 2018.

\bibitem{ozdemir20193d}
O.~Ozdemir, R.~L. Russell, and A.~A. Berlin, ``A 3d probabilistic deep learning
  system for detection and diagnosis of lung cancer using low-dose ct scans,''
  \emph{IEEE transactions on medical imaging}, vol.~39, no.~5, pp. 1419--1429,
  2019.

\bibitem{ardila2019end}
D.~Ardila, A.~P. Kiraly, S.~Bharadwaj, B.~Choi, J.~J. Reicher, L.~Peng, D.~Tse,
  M.~Etemadi, W.~Ye, G.~Corrado \emph{et~al.}, ``End-to-end lung cancer
  screening with three-dimensional deep learning on low-dose chest computed
  tomography,'' \emph{Nature medicine}, vol.~25, no.~6, pp. 954--961, 2019.

\bibitem{zhu2018deeplung}
W.~Zhu, C.~Liu, W.~Fan, and X.~Xie, ``Deeplung: Deep 3d dual path nets for
  automated pulmonary nodule detection and classification,'' in \emph{2018 IEEE
  Winter Conference on Applications of Computer Vision (WACV)}.\hskip 1em plus
  0.5em minus 0.4em\relax IEEE, 2018, pp. 673--681.

\bibitem{ren2016faster}
S.~Ren, K.~He, R.~Girshick, and J.~Sun, ``Faster r-cnn: towards real-time
  object detection with region proposal networks,'' \emph{IEEE transactions on
  pattern analysis and machine intelligence}, vol.~39, no.~6, pp. 1137--1149,
  2016.

\bibitem{szegedy2015going}
C.~Szegedy, W.~Liu, Y.~Jia, P.~Sermanet, S.~Reed, D.~Anguelov, D.~Erhan,
  V.~Vanhoucke, and A.~Rabinovich, ``Going deeper with convolutions,'' in
  \emph{Proceedings of the IEEE conference on computer vision and pattern
  recognition}, 2015, pp. 1--9.

\bibitem{xu2019deep}
Y.~Xu, A.~Hosny, R.~Zeleznik, C.~Parmar, T.~Coroller, I.~Franco, R.~H. Mak, and
  H.~J. Aerts, ``Deep learning predicts lung cancer treatment response from
  serial medical imaging,'' \emph{Clinical Cancer Research}, vol.~25, no.~11,
  pp. 3266--3275, 2019.

\bibitem{horst2020delivering}
C.~Horst, J.~L. Dickson, S.~Tisi, M.~Ruparel, A.~Nair, A.~Devaraj, and S.~M.
  Janes, ``Delivering low-dose ct screening for lung cancer: a pragmatic
  approach,'' \emph{Thorax}, vol.~75, no.~10, pp. 831--832, 2020.

\bibitem{jenkinson2001global}
M.~Jenkinson and S.~Smith, ``A global optimisation method for robust affine
  registration of brain images,'' \emph{Medical image analysis}, vol.~5, no.~2,
  pp. 143--156, 2001.

\bibitem{jenkinson2002improved}
M.~Jenkinson, P.~Bannister, M.~Brady, and S.~Smith, ``Improved optimization for
  the robust and accurate linear registration and motion correction of brain
  images,'' \emph{Neuroimage}, vol.~17, no.~2, pp. 825--841, 2002.

\bibitem{ronneberger2015u}
O.~Ronneberger, P.~Fischer, and T.~Brox, ``U-net: Convolutional networks for
  biomedical image segmentation,'' in \emph{International Conference on Medical
  image computing and computer-assisted intervention}.\hskip 1em plus 0.5em
  minus 0.4em\relax Springer, 2015, pp. 234--241.

\bibitem{he2016deep}
K.~He, X.~Zhang, S.~Ren, and J.~Sun, ``Deep residual learning for image
  recognition,'' in \emph{Proceedings of the IEEE conference on computer vision
  and pattern recognition}, 2016, pp. 770--778.

\bibitem{hu2018squeeze}
J.~Hu, L.~Shen, and G.~Sun, ``Squeeze-and-excitation networks,'' in
  \emph{Proceedings of the IEEE conference on computer vision and pattern
  recognition}, 2018, pp. 7132--7141.

\bibitem{py06nimg}
P.~A. Yushkevich, J.~Piven, H.~Cody~Hazlett, R.~Gimpel~Smith, S.~Ho, J.~C. Gee,
  and G.~Gerig, ``User-guided {3D} active contour segmentation of anatomical
  structures: Significantly improved efficiency and reliability,''
  \emph{Neuroimage}, vol.~31, no.~3, pp. 1116--1128, 2006.

\bibitem{lin2017focal}
T.-Y. Lin, P.~Goyal, R.~Girshick, K.~He, and P.~Doll{\'a}r, ``Focal loss for
  dense object detection,'' in \emph{Proceedings of the IEEE international
  conference on computer vision}, 2017, pp. 2980--2988.

\bibitem{NEURIPS2019_9015}
\BIBentryALTinterwordspacing
A.~Paszke, S.~Gross, F.~Massa, A.~Lerer, J.~Bradbury, G.~Chanan, T.~Killeen,
  Z.~Lin, N.~Gimelshein, L.~Antiga, A.~Desmaison, A.~Kopf, E.~Yang, Z.~DeVito,
  M.~Raison, A.~Tejani, S.~Chilamkurthy, B.~Steiner, L.~Fang, J.~Bai, and
  S.~Chintala, ``Pytorch: An imperative style, high-performance deep learning
  library,'' in \emph{Advances in Neural Information Processing Systems 32},
  H.~Wallach, H.~Larochelle, A.~Beygelzimer, F.~d\textquotesingle
  Alch\'{e}-Buc, E.~Fox, and R.~Garnett, Eds.\hskip 1em plus 0.5em minus
  0.4em\relax Curran Associates, Inc., 2019, pp. 8024--8035. [Online].
  Available:
  \url{http://papers.neurips.cc/paper/9015-pytorch-an-imperative-style-high-performance-deep-learning-library.pdf}
\BIBentrySTDinterwordspacing

\bibitem{tammemagi2013selection}
M.~C. Tammem{\"a}gi, H.~A. Katki, W.~G. Hocking, T.~R. Church, N.~Caporaso,
  P.~A. Kvale, A.~K. Chaturvedi, G.~A. Silvestri, T.~L. Riley, J.~Commins
  \emph{et~al.}, ``Selection criteria for lung-cancer screening,'' \emph{New
  England Journal of Medicine}, vol. 368, no.~8, pp. 728--736, 2013.

\bibitem{cristianini2000introduction}
N.~Cristianini, J.~Shawe-Taylor \emph{et~al.}, \emph{An introduction to support
  vector machines and other kernel-based learning methods}.\hskip 1em plus
  0.5em minus 0.4em\relax Cambridge university press, 2000.

\bibitem{liaw2002classification}
A.~Liaw, M.~Wiener \emph{et~al.}, ``Classification and regression by
  randomforest,'' \emph{R news}, vol.~2, no.~3, pp. 18--22, 2002.

\end{thebibliography}

\end{document}


\title{Supplementary Information}
\author{SUMMIT consortium refers to all co-authors that comprise the SUMMIT study group including:

Sam M Janes$^{1}$, Jennifer L Dickson$^{1}$, Carolyn Horst$^{1}$, Sophie Tisi$^{1}$, Helen Hall$^{1}$, Priyam Verghese$^{1}$, Andrew Creamer$^{1}$, Thomas Callender$^{1}$, Ruth Prendecki$^{1}$, Amyn Bhamani$^{1}$, Mamta Ruparel$^{1}$, Allan Hackshaw$^{2}$, Laura Farrelly$^{2}$, Jon Teague$^{2}$, Anne-Marie Mullin$^{2}$, Kitty Chan$^{2}$, Rachael Sarpong$^{2}$, Malavika Suresh$^{2}$, Samantha L Quaife$^{3}$, Arjun Nair$^{4}$, Anand Devaraj$^{5,6}$, Kylie Gyertson$^{4}$, Vicky Bowyer$^{4}$, Ethaar El-Emir$^{4}$, Judy Airebamen$^{4}$, Alice Cotton$^{4}$, Kaylene Phua$^{4}$, Elodie Murali$^{4}$, Simranjit Mehta$^{4}$, Janine Zylstra$^{4}$, Karen Parry-Billings$^{4}$, Columbus Ife$^{4}$, April Neville$^{4}$, Paul Robinson$^{4}$, Laura Green$^{4}$, Zahra Hanif$^{4}$, Helen Kiconco$^{4}$, Ricardo McEwen$^{4}$, Dominique Arancon$^{4}$, Nicholas Beech$^{4}$, Derya Ovayolu$^{4}$, Christine Hosein$^{4}$, Sylvia Patricia Enes$^{4}$, Qin April Neville$^{4}$, Jane Rowlands$^{4}$, Aashna Samson$^{4}$, Urja Patel$^{4}$, Fahmida Hoque$^{4}$, Hina Pervez$^{4}$, Sofia Nnorom$^{4}$, Moksud Miah$^{4}$, Julian McKee$^{4}$, Mark Clark$^{4}$, Jeannie Eng$^{4}$, Fanta Bojang$^{4}$, Claire Levermore$^{4}$, Anant Patel$^{7}$, Sara Lock$^{8}$, Rajesh Banka$^{9}$, Angshu Bhowmik$^{10}$, Ugo Ekeowa$^{11}$, Zaheer Mangera$^{12}$, William M Ricketts$^{13}$, Neal Navani$^{4}$, Terry O'Shaughnessy$^{13}$, Charlotte Cash$^{7}$, Magali Taylor$^{4}$, Samanjit Hare$^{7}$, Tunku Aziz$^{13}$, Stephen Ellis$^{13}$, Anthony Edey$^{14}$, Graham Robinson$^{15}$, Alberto Villanueva$^{16}$, Hasti Robbie$^{17}$, Elena Stefan$^{18}$, Charlie Sayer$^{19}$, Nick Screaton$^{20}$, Navinah Nundlall$^{21}$, Lyndsey Gallagher$^{4}$, Andrew Crossingham$^{4}$, Thea Buchan$^{4}$, Tanita Limani$^{4}$, Kate Gowers$^{1}$, Kate Davies$^{1}$, John McCabe$^{1}$, Joseph Jacob$^{1,21}$, Karen Sennett$^{22}$, Tania Anastasiadis$^{23}$, Andrew Perugia$^{24}$, James Rusius$^{24}$.

1	Lungs For Living Research Centre, UCL Respiratory, University College London, London

2	CRUK and UCL Cancer Trials Centre, University College London, London

3	Centre for Prevention, Detection and Diagnosis, Wolfson Institute of Population Health, Barts and The London School of Medicine and Dentistry, Queen Mary University of London, London

4	University College London Hospitals NHS Foundation Trust, London

5	Royal Brompton and Harefield NHS Foundation Trust, London

6	National Heart and Lung Institute, Imperial College, London

7	Royal Free London NHS Foundation Trust, London

8	Whittington Health NHS Trust, London

9	Barking, Havering and Redbridge University Hospitals NHS Trust, Essex

10	Homerton University Hospital Foundation Trust, London

11	The Princess Alexandra Hospital NHS Trust, Essex

12	North Middlesex University Hospital NHS Trust, London

13	Barts Health NHS Trust, London

14	North Bristol NHS Trust, Bristol

15	Royal United Hospitals Bath NHS Foundation Trust, Bath

16	Surrey and Sussex Healthcare NHS Trust, Surrey

17	King's College Hospital NHS Foundation Trust, London

18	The Princess Alexandra Hospital NHS Trust, London

19	University Hospitals Sussex NHS Foundation Trust, Sussex

20	Royal Papworth Hospital NHS Foundation Trust, Cambridge

21	Centre for Medical Image Computing (CMIC), London

22	Killick Street Health Centre, London

23	Tower Hamlets Clinical Commissioning Group, London

24	Noclor Research Support, London}
%
%
%
\markboth{Journal of \LaTeX\ Class Files}%
{Shell \MakeLowercase{\textit{et al.}}:}
%
%
%
\maketitle
\thispagestyle{empty}
\pdfoutput=1
\ifCLASSOPTIONcaptionsoff
  \newpage
\fi